# A Magnetic Spectrometer for Electron-Positron Pair Spectroscopy in Storage Rings

S. Hagmann[1], P.M. Hillenbrand[1, 2], Yu. A. Litvinov[1], U. Spillmann[1], Th. Stöhlker[1,3,4]

[1]GSI Helmholtz-Zentrum Darmstadt, [2]Columbia University, N.Y., USA, [3]Helmholtz Institut Jena, [4]IOQ, Universität Jena and HI Jena

*We report an analysis of electron-optical properties of a toroidal magnetic sector spectrometer and examine parameters for its implementation in a relativistic heavy-ion storage ring like HESR at the future FAIR facility. For studies of free-free pair production in heavy- ion atom collisions, this spectrometer exhibits very high efficiencies for coincident $e^+$-$e^-$ pair spectroscopy over a wide range of momenta of emitted lepton pairs. The high coincidence efficiency of the spectrometer is the key for stringent tests of theoretical predictions for the phase space correlation of lepton vector momenta in free-free pair production.*

Electron–positron pair production has evolved into a central topic of QED in extreme fields as the coupling between the lepton field and the electromagnetic field is close to one [1-4]. Most recently the surprisingly high cross sections observed for pair production in relativistic heavy-ion atom collisions [1] have aroused new interest in this process even in the accelerator design community as capture from pair production has been identified as a critical, potentially luminosity limiting process in relativistic heavy ion colliders. The enormous cross sections observed (~3b) can be traced [1] to the large transverse electric fields $E_{transv} \sim \gamma$. Theory predicts a scaling of the total cross section for free-free pair production [1, 3] $\sigma_{free-free} \sim (Z_{proj})^2 (Z_{tar})^2 \ln^3 \gamma$ and a very complex relation between the angular emission patterns of electron and positron [4]. The deep relation between pair production and the short-wavelength-limit (SWL) of electron nucleus bremsstrahlung (eNBS) and the need for corroborating experiments has been emphasized strongly by theory [5, 6]. The future relativistic storage ring HESR at FAIR [7,9] with a collision energy range up to $\gamma \cong 6$ will be best suited to study for ion-induced pair-production all channels which may be distinguished experimentally [3] and the SWL of eNBS as well. A corresponding future experimental investigation of the dynamics of the heavy-ion induced free-free pair creation up to

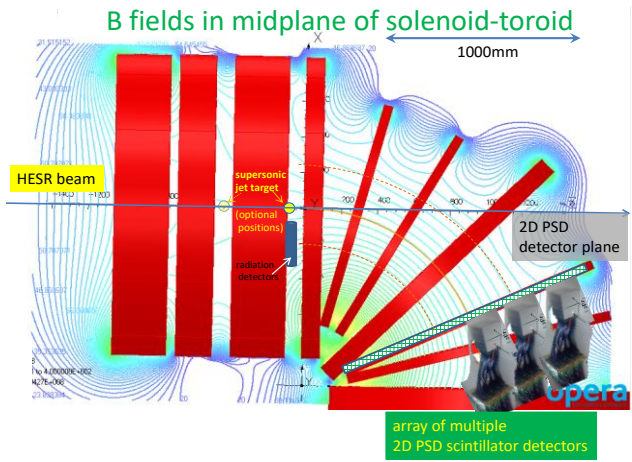

**Fig. 1.** *Cut through midplane of coil assembly of a toroidal magnetic lepton spectrometer for lepton energies up to 20 MeV; detectors are positioned at coil S8.*

$\gamma \sim 6$ following

$$X^{Z+} + A \rightarrow X^{Z+} + \{A^*\} + e^+(p_+, \theta_+) + e^-(p_-, \theta_-) \qquad (1)$$

is the motivation for the current spectrometer design study: a coincident detection of both outgoing leptons and their vector momenta as is constitutive in a complete description of free-free pair production, is not practically possible with traditional dispersive instruments covering only a small solid angle for each lepton.

We have therefore employed OPERA-3D [8] to study the electro-optical properties of a magnetic toroidal spectrometer with very large effective solid angle. It will enable coincident detection of the vector momenta of electrons and positrons from a free-free pair (see fig. 1) to be implemented in the target-zone of the HESR[7,9]. In a toroidal B-field e$^-$ and e$^+$ are momentum-dispersed perpendicular to the bend plane of the toroid in opposite directions. For ion-atom collisions, the open phase space spanned by the two leptons of a free-free pair in the continuum is very large and is practically not constrained by any kinematic conservation rules. Theory [3] has predicted that the two leptons of a pair may be indeed emitted with a nearly unrestricted, very wide range of difference angles (in the emitter frame). We illustrate in fig. 2 for typical lepton energies in the emitter frame for 5.1 AGeV Au + Au, i.e. 200 keV to 2 MeV, the kinematic relations and the very large phase space for both leptons resulting in the laboratory for emission of leptons over a wide range of kinetic energies. In the absence of kinematic restrictions for coincident lepton pairs, a unique attribution of the primordial vector momentum of the lepton in the emitter frame (the emitter frame characterized by $\beta_0$ and $\gamma_0$, a lepton in the emitter frame by p', β', γ') using

$$(p'c)^2 = (pc)^2 \sin^2\theta_{lab} + \gamma_0^2\left(pc\cos\theta_{lab} - \beta_0\sqrt{(pc)^2 + m^2c^4}\right)^2 \quad (2)$$

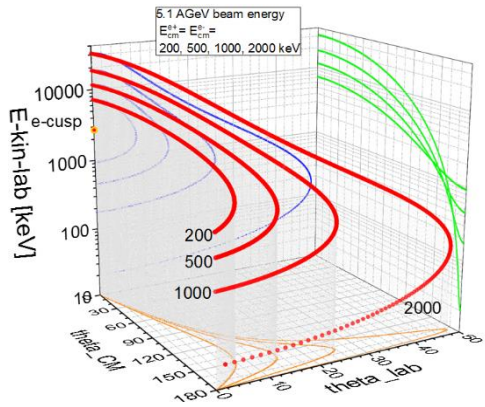

can only be provided when both, the lepton's laboratory momentum p/ energy E$_{lab}$ and the observation angle θ$_{lab}$, are simultaneously determined, as follows from fig. 2. The corresponding equation relating laboratory frame and emitter frame angles is

*Fig. 2. Kinematical relations for leptons emitted by a fast projectile, for lepton energies 200 keV to 2 MeV in the emitter frame. Laboratory kinetic energies of the leptons extend from ≈10 keV to beyond 30 MeV in the forward direction [10].*

$$\cos\theta' = \frac{-\frac{\beta_0}{\beta'}\gamma_0^2 \sin^2\theta_{lab} \pm \sqrt{\gamma_0^2(1-(\frac{\beta_0}{\beta'})^2)\sin^2\theta_{lab} + \cos^2\theta_{lab}}\;\cos\theta_{lab}}{\gamma_0^2 \sin^2\theta_{lab} + \cos^2\theta_{lab}} \quad . \quad (3)$$

For projectiles in the HESR with a specific energy 5.1 AGeV the kinematics for leptons observed with e.g. E$_{lab}$=6000 keV laboratory kinetic energy does constrain resulting emitter frame kinetic energy range only from 200 keV to around 2000 keV while the corresponding emitter frame angles may range between 60$^0$ and 120$^0$. Fig. 2 also shows that the large laboratory observation angles correspond to small laboratory energies for backward emission in the emitter frame θ$_{CM}$≥150$^0$ facilitating detection. Leptons in a toroidal magnetic field will execute near-helical motions. We find that for a given B-field one can tune lepton momenta $\beta_0\gamma_0/n$, $n\geq 1$, such that all trajectories independent of their forward laboratory emission angle will intersect in a node $X_n$ on the detector plane (fig. 3). For all other lepton momenta $\beta\gamma$ the parallel and perpendicular momenta components can be determined by intersections on 2D-PSD detectors. Instead of directly measuring $\Theta_{lab}$ of a lepton, however, it is far more convenient to experimentally determine its gyroradius $r_g=(\beta\gamma)_\perp m_0 c/qB$ in the toroidal B-field and subsequently calculate its laboratory transverse momentum. Due to the apperance of nodes on the 2D-PSD detector for certain lepton momenta, the relation between the true gyroradius $r_g$ and the effective $r_{g\text{-eff}}$ obtained from the location of the intercept in the detector plane must be, however, established in the spectrometer calibration [10] using conversion electron lines, because only for lepton trajectories intersecting the detector plane outside the nodes $X_n$ one may eventually obtain $r_g$ and the transverse momentum and its azimuth from $r_{g\text{-eff}}$.(fig. 4) The first step of the experimental momentum/energy calibration establishes the arc $C_-$ (and mirrored on center plane $C_+$) using the narrow electron cusp ($\theta_{lab}=0^0 \leftrightarrow p_\perp(lab)=0$ to a very good approximation) at varying beam energies with electron cusp momenta $p_{cusp}=(\beta\gamma)_{beam}$ and additionally conversion electron lines from radioactive sources equipped with an aperture for $\theta_{lab}=0^0$. It is a useful property of the arcs that their geometry only depends on the toroidal angle and radius but is independent of the magnetic field; with changing B-field a node for a given lepton momentum slides along the invariant arcs ( fig 5).

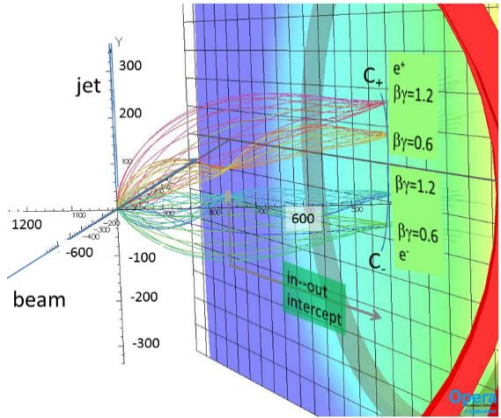

**Fig. 3.** *Vertical dispersion of leptons in the toroidal field; nodes lie on arcs $C_+$ and $C_-$.*

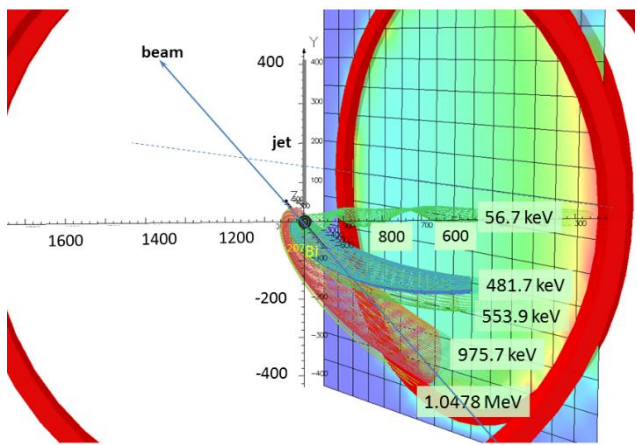

**Fig. 4**. *Energy/momentum calibration of the toroidal magnetic spectrometer using Pb KLL Auger and conversion electron lines emitted by a $^{207}$Bi source and detected at $66^0$ for a B-field setting of 108.7 G. Besides the Pb KLL Auger group we show in this illustration only the strongest components of the conversion electron line groups; besides the strong 481.7 keV line the 553.9 keV line with $\approx$ 25% intensity of the former and the 975.7 keV line with the 1047.8 keV line which also has around a quarter the intensity of the former.*

For the general gyro-radius/momentum calibration, i.e. intercepts in the detector plane off the arc $C_-$, it is convenient to use angle defining apertures, e.g. $0^0$, $5^0$ and $15^0$ with respect to the beam axis, for a $^{207}$Bi conversion electron source over a range of different magnetic B-fields in the toroid. In fig. 5 we illustrate that for a conversion electron energy given, e.g. 975 keV, the B-field may either be tuned to focus all electrons onto one focal spot in the detector plane, as here shown for 234 G, or to any lower value, e.g. here 165 G, where the effective gyro-radius on the detector plane can be related to the transverse electron momentum via the known laboratory emission angle in the calibration source mask.

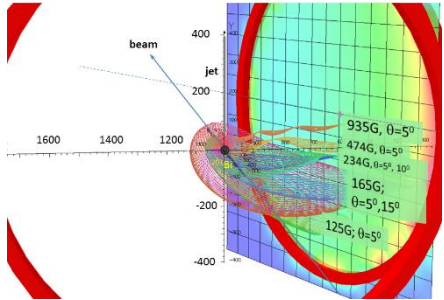

**Fig. 5**. *Mapping of the 975.7 keV conversion electron line from $^{207}$Bi onto the detector plane at $66^0$ and toroidal magnetic B-field for fields ranging from 125 G top 935 G. The source located in the target zone beneath the supersonic jet target is in this OPERA simulation equipped with angle defining apertures to better illustrate the mapping used for calibration.*

The combination of ≈1% energy resolution and 2D position sensitive detection of the electron accomplishes the unambiguous identification where electrons of different energies may hit the same location on the detector. This permits to calibrate the apparent gyroradius $r_{g\text{-eff}}$ for the respective transverse momentum component (and the corresponding true gyro-radius $r_g$) selected by the angle setting in the aperture, e.g. of the prevailing 975 keV conversion electron line over a range of magnetic fields [10]. In fig. 5 we show how the 975keV conversion electron line for a range of B-fields slides along the arc C-. For 165 G we illustrate how using apertures for two different angles the procedure for momentum calibration of the locations of the intercept in the detector plane is accomplished. It can also be nicely seen that for 234 G all electrons from the 975.7 keV line emerging in the forward direction with $\theta \leq 5^0$ converge on a node after $66^0$ on the arc C-, i.e. at coil 8 of the toroid.The coincident double, triple and quadruple differential cross sections for free-free lepton pair productions may thus be determined for the first time over the entire phase space by coincident detection of leptons in the detector plane for a choice of a few appropriate B-field settings.

**Acknowledgments**

Within this project Yu. A. Litvinov has received funding from the European Research Council (ERC) under the European Union's Horizon 2020 research and innovation programme (grant agreement No 682841 "ASTRUm'')
 P.M. Hillenbrand acknowledges support by DFG project HI 2009/1-1.